\documentclass[aps,prb,twocolumn,showpacs,amsmath,amssymb]{revtex4-1}

\usepackage{graphicx}
\usepackage{dcolumn}
\usepackage{bm}
\usepackage{subfigure}

\begin{document}

\title{Atomic Force Microscope manipulation of Ag atom on the Si(111) surface}

\author{Batnyam Enkhtaivan}
\author{Atsushi Oshiyama}

\affiliation{Department of Applied Physics, The University of Tokyo, Hongo, Tokyo 113-8656, Japan}

\date{\today}

\begin{abstract}
We present first-principles total-energy electronic-structure calculations that provide the microscopic mechanism of the Ag atom diffusion between the half unit cells (HUCs) on the Si(111)-(7$\times$7) surface with and without the tip of the atomic force microscope (AFM). We find that, without the presence of the AFM tip, the diffusions between the 
two HUCs are almost symmetric with the energy barrier of about 1 eV in the both directions. The diffusion is a two-step process with an intermediate metastable configuration in which the Ag atom is at the boundary of the HUCs. With the presence of the tip, we find that the reaction pathways are essentially the same, but the energy barrier in one direction is substantially reduced to be 0.2 - 0.4 eV by the tip, while that of the diffusion in the reverse direction remains larger than 1 eV. The Si tip reduces the energy barrier more than the Pt tip due to the flexibility of the tip apex structure. In addition to the reduction of the barrier, the tip traps the diffusing adatom preventing the diffusion in the reverse direction. Also we find that the shape of the tip apex structure is important for the trapping ability of the adatom. When the tip apex structure is blunt, the adatom interacts with the tip atom other than the tip apex atom. The bond formation between the AFM tip atom and the surface adatom is essential for the atom manipulation using the AFM tip. Our results show that the atom manipulation is possible with both the metallic and semiconducting AFM tips.
\end{abstract}

\maketitle

\section{\label{sec:introduction}INTRODUCTION}

Atom-scale resolved structural images of solid surfaces have been achieved by the scanning tunneling microscope (STM)\cite{stm} and the atomic force microscope (AFM)\cite{afm} by measuring the tunneling electric current between the probing tip and the surface, and the force acting on the probing tip, respectively. Apart from the surface imaging, these scanning probe techniques have also been utilized for manipulating atoms on surfaces. Eigler and Schweizer demonstrated the first atom manipulation by creating the IBM company logo with Xe atom on Ni surface by laterally moving the atoms by the STM.\cite{eigler} In the following year, Lyo and Avouris reported removal and deposition of single Si atom on Si surface, showing the possibility of the vertical manipulation also using the STM.\cite{lyo} A decade later, Oyabu \textit{et. al.,} reported atom manipulation on Si surface by the AFM,\cite{oyabu1} expanding the feasibility of atom manipulation to non-conductive surfaces.

Since then, many types of atom manipulation by the AFM probe has been achieved. They include vertical interchange of the tip and surface atoms on the Sn-covered Si(111) surface,\cite{vertical_interchange} lateral interchange of adatoms on the Ge(111)\cite{lateral_interchange1} and on the Si(111)\cite{lateral_interchange2} surfaces, lateral manipulation of single Si and Ge adatoms on the Si(111)\cite{lateral_shift1,lateral_shift2} and the Ge(111)\cite{lateral_shift3} surfaces, respectively.

Recently, Sugimoto {\it et al.} controlled the diffusion of Ag, Au, Pb, Sn and Si atoms by STM/AFM probe on the Si(111)-(7$\times$7) surface.\cite{gate} Here, STM/AFM is a combined equipment of STM and AFM which simultaneously measures the tunneling current between the tip and surface, and frequency shift of the oscillating cantilever with Pt-Ir coated Si tip. It was observed that the adatoms diffuse in the surface half unit cell (HUC), while the inter-HUC diffusion of adatom is hindered by high potential energy barrier. Further, Sugimoto {\it et al.} induced the inter-HUC diffusion of adatom by approaching the probe of STM/AFM or AFM to the boundary of HUCs slightly off to the one side. By analyzing the simultaneously measured tunneling current and the frequency shift of the cantilever, they concluded that the diffusion controlling is purely due to the chemical interaction between the tip and the diffusing atom; not due to the electric field and current. To support this, they also performed the controlling of Sn diffusion on the Si(111) surface by the AFM with the Si tip. This opened exciting possibilities of the fabrication of atom-number determined nanoclusters and studying their properties.\cite{catalyst1,catalyst2,catalyst3} Albeit the experimental demonstrations, the microscopic mechanisms of the gate controlling is unclear for no theoretical study is reported.

The growth of silver on the Si(111) surface has been foci of many studies,\cite{silver1,silver2,silver3,silver4} because of the low intermixing between the two substances. Also, many studies in both experiment and theory regarding the state of the single Ag atom on the Si(111)-(7$\times$7) have been done.\cite{agondas1exp,agondas2exp,agondas3exp,agondas4exp,agondastheory,ihucbarrier1,ihucbarrier2} The theoretical works have concentrated on clarification of the stable positions and diffusion pathway inside the HUC.\cite{pes,agondastheory} Wang \textit{et. al.,} presented the static potential energy surface (PES) of single Ag atom on the Si(111) surface, and gave clear understanding of the typical STM images.\cite{pes}

Therefore, to clarify the microscopic mechanism of the diffusion controlling of adatom by the tip of the scanning probe microscope, we choose the system in which Ag atom adsorbed on the Si(111) substrate as a representative system and have performed extensive density-functional calculations. We first obtain the PES of Ag atom confirming the result of previous work\cite{pes} and examine the inter-HUC diffusion without the AFM tip. We identified two diffusion pathways: There is a metastable atomic configuration in the middle of each pathway and two-step diffusion takes place. The corresponding energy barriers are calculated. We examine the modification of the diffusion process by the Si and the Pt tips, and find that the presence of the tip lowers the diffusion barrier, enhancing the inter-HUC diffusion, and traps the Ag atom near it. The extent of the modification significantly varies depending on the flexibility of the tip apex structure.

The paper is organized as follows. The calculational methods and the pertinent conditions for the calculations are explained in Sec. \ref{sec:method}. The inter-HUC diffusion of the Ag atom without the AFM tip is described in Sec.~\ref{subsec:notip}. In Sec.~\ref{subsubsec:si_tip} and Sec.~\ref{subsubsec:pt_tip}, the modification of the adatom diffusion process by the Si and the Pt tips is described, respectively. Finally, we summarize our findings in Sec.~\ref{sec:summary}.

\section{\label{sec:method}CALCULATIONS}

Calculations are performed within the framework of the density functional theory (DFT)\cite{dft1,dft2} using the Vienna Ab initio Simulation Package (VASP).\cite{vasp1,vasp2} The generalized gradient approximation (GGA)\cite{gga} is used for the calculation of the exchange-correlation energy. Projector augmented-wave (PAW) potentials \cite{paw} are adopted to describe the electron-ion interaction. We use the cutoff energy of 250 eV for the plane-wave basis.

Each substrate surface is simulated by a repeating slab model. When the AFM tip is included, the atomic slab is separated from its adjacent image slabs by the vacuum region so that the atomic distances between the different slabs are more than 6 \text{\AA}, which is found to be large enough to neglect the interaction between the slab and its images.\cite{bato} Without the AFM tip, the slabs are separated by more than 8 \text{\AA} distance. The slab for the Si(111)-(7$\times$7) surfaces is simulated by six atomic layers in addition to the adatom layer. The atoms at the bottommost layer of the slab are terminated with H atoms to remove unsuitable dangling bonds electronically. 
In the lateral directions,  Si(111)-(7$\times$7) surface is simulated by the dimer-adatom-stacking-fault (DAS) model of Takayanagi \textit{et al.}.\cite{das} Only $\Gamma$ point is sampled for the Brillouin zone integration for the supercell cells are large. Structural optimization is performed using calculated Hellmann-Feynman forces. All the atoms except for the bottommost layer atoms and the attached H atoms are relaxed until the forces acting on the atoms are smaller than 0.1 eV/\text{\AA}. The conditions explained above assure that the numerical error of the binding energy of Ag atom is less than 5 meV. The binding energy is the total energy difference of the isolated Ag atom and the isolated surface from the combined system of the Ag atom and the surface.

The PES is obtained by calculating the binding energy of the Ag atom at the grid points of 64$\times$64 rectangular grids in the rectangular unit cell of the Si(111)-c(7$\times$7) surface and by interpolating the values between the grid points with Fourier transformation. In these calculations, the (x,y) coordinates of the Ag atom are fixed.

To identify the reaction pathways of the diffusing atom and the corresponding energy barriers, we adopt climbing-image nudged elastic band (CINEB) method.\cite{cineb} This method provides the highest saddle point energy, while partly assuring the continuity of the reaction pathway compared with the hyperplane constraint method,\cite{jeong} by introducing fictitious elastic forces during the energy minimization. In each barrier calculation of an elementary reaction, 3 image configurations are considered between the initial and the final states.

\begin{figure}
\centering
\includegraphics[width=0.9\linewidth]{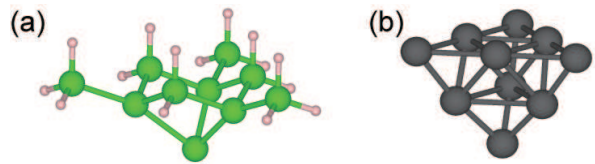}
\caption{
(Color online) The model of the (a) Si and (b) Pi tips considered in the calculation. 
The green (medium), purple (small), and black (large) spheres depict Si, H, and Pt atoms, respectively. 
}
\label{fig:tip}
\end{figure}

We consider semiconductor and metallic AFM tips composed of Si and Pt atoms, respectively. The Pt tip is utilized as simple version of the Pt-Ir coated tip used in the experiment.\cite{gate} The tips are simulated by the atomistic model shown in Fig.~\ref{fig:tip}. The Si tip model consists of 10 Si atoms and 15 H atoms. This model is used in previous works.\cite{afm_dft_perez1,afm_dft_perez2,lateral_shift2,lateral_shift3,jarvis_moriarty,tip_apex_shape,bato} The Pt tip model consists of 10 Pt atom. In our calculations, we have done structural minimization of these tips along with the surface atomic configurations. The H atoms and the Si atoms bonding with the H atoms in the Si tip, and the bottom-face Pt atoms in the Pt tip, however, are fixed during the geometry optimization.

The force acting on the AFM tip is obtained by summing force acting along z-direction on the fixed atoms of the tip.

\section{\label{sec:results}RESULTS}

\subsection{\label{subsec:notip} Adatom diffusion without the AFM tip}

\begin{figure}
\centering
\includegraphics[width=0.95\linewidth]{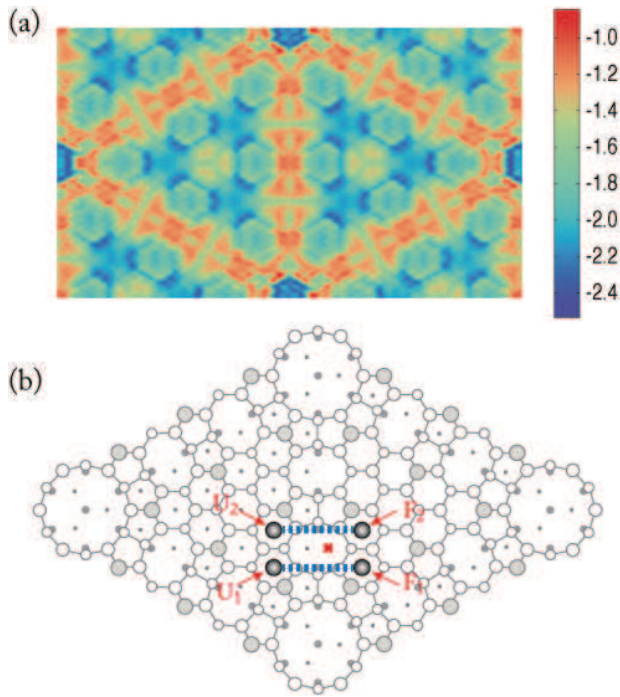}
\caption{
(Color online) (a) The calculated potential energy surface (PES) of Ag adatom on the Si(111)-(7$\times$7) surface, and (b) the schematic top view of the Si(111)-(7$\times$7) surface.
In the colorbar of (a), the blue color shows large binding energy and the red color shows smaller binding energy. The unit is eV. In (b), the distinct stable and metastable positions of Ag atom are labeled as $U_1$ and $U_2$ in the UHUC, and $F_1$ and $F_2$ in the FHUC. The diffusion pathways between these positions are shown by the blue dashed lines. The red x-mark shows the lateral position of the tip apex atom of the AFM tip.}
\label{fig:pes}
\end{figure}

The calculated PES of Ag atom on the Si(111)-(7$\times$7) surface is shown in Fig.~\ref{fig:pes} (a). There are 3 potential energy wells, the areas in blue color, in each basin\cite{basin} and the depths of the potential wells are asymmetric between the HUCs. The distinct (meta)stable adsorption sites are labeled as $F_1$ and $F_2$ in the faulted-HUC (FHUC), and $U_1$ and $U_2$ in the unfaulted-HUC (UHUC) [see Fig.~\ref{fig:pes} (b)]. The $U_1$, $U_2$, and $F_2$ sites have the binding energy of about 2.30 eV, and the $F_1$ site has about 30 meV larger binding energy than the others. These results are in agreement with the previous works.\cite{pes,agondastheory}

\begin{figure}
\centering
\includegraphics[width=0.7\linewidth]{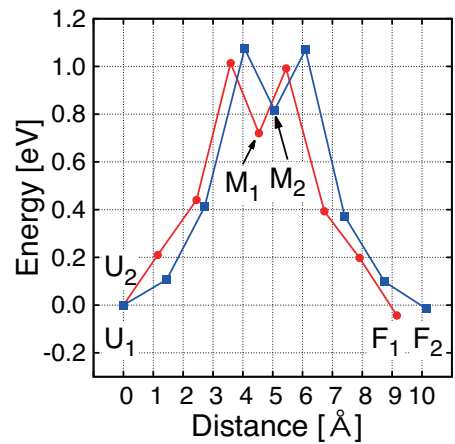}
\caption{
(Color online) The energy profiles of the Ag adatom diffusion along the \textit{pathway 1} and the \textit{pathway 2} without the AFM tip. The energy profiles between $U_1$ and $F_1$, and between $U_2$ and $F_2$ are shown in red (circle) and blue (square) graphs, respectively. The x axis is the distance between Ag atoms in the NEB image configurations and the initial structure. The y axis is the total energy difference of the NEB image configurations from the initial structure. 
}
\label{fig:notip}
\end{figure}

From the PES, it is clear that the diffusion between the HUCs occurs between $U_1$ and $F_1$ (\textit{pathway 1}), and between $U_2$ and $F_2$ (\textit{pathway 2}). The pathways are shown by blue dashed lines in Fig.~\ref{fig:pes} (b). The obtained energy profiles along them are shown in Fig.~\ref{fig:notip}. There are metastable configurations right in the middle of the \textit{pathway 1} and the \textit{pathway 2}, each labeled as $M_1$ and $M_2$, respectively. The energy barriers of $U_1 \rightarrow M_1$ and $M_1 \rightarrow F_1$ processes are 1.01 and 0.27 eV, respectively, and those of the reverse $F_1 \rightarrow M_1$ and $M_1 \rightarrow U_1$ processes are 1.03 and 0.29 eV, respectively. The high energy barriers for the diffusion in either direction between the HUCs show that the inter-HUC diffusion is infrequent event at room temperature. There is a slight asymmetry in the diffusion along the \textit{pathway 1} (see Fig.~\ref{fig:notip}). On the contrary to this, the diffusion along the \textit{pathway 2} is almost symmetric, having diffusion barriers of 1.07 and 0.25 eV for $U_2 \rightarrow M_2 \rightarrow F_2$ process, and 1.08 and 0.26 eV for $F_2 \rightarrow M_2 \rightarrow U_2$ process. The experimentally observed energy barriers of $F_1 \rightarrow U_1$ and $U_1 \rightarrow F_1$ processes are about 0.81 and 0.9 eV, respectively.\cite{ihucbarrier1} Compared to the experimentally observed diffusion barriers, these energy barriers are slightly overestimated by about 0.1 - 0.2 eV and the difference in the energy barriers of the diffusion into and out of UHUC diffusions is smaller. Considering the approximations used in the calculation, the calculated energy barriers are in well agreement with the experiment.

\subsection{\label{subsec:tip} Ag atom diffusion with the AFM tip}

In this section, we show the modifications of the diffusion barriers and pathways by the AFM tips (Si and Pt tips). We have considered the cases in which the tip-surface distances are 3.5, 4.0, 4.5, and 5.0 \text{\AA}. The tip-surface distance is defined as the distance between the tip apex atom and the surface adatom of the Si(111) surface before the structural relaxation. For simplicity, we focus on the UHUC $\rightarrow$ FHUC diffusion of the Ag atom. The AFM tip is placed at the red x-mark as shown in Fig.~\ref{fig:pes}, corresponding to the tip position in the experiment.\cite{gate}

\subsubsection{\label{subsubsec:si_tip} The effect of the Si tip on the diffusion}

\begin{figure*}[htbp]
\centering
\includegraphics[width=0.95\linewidth]{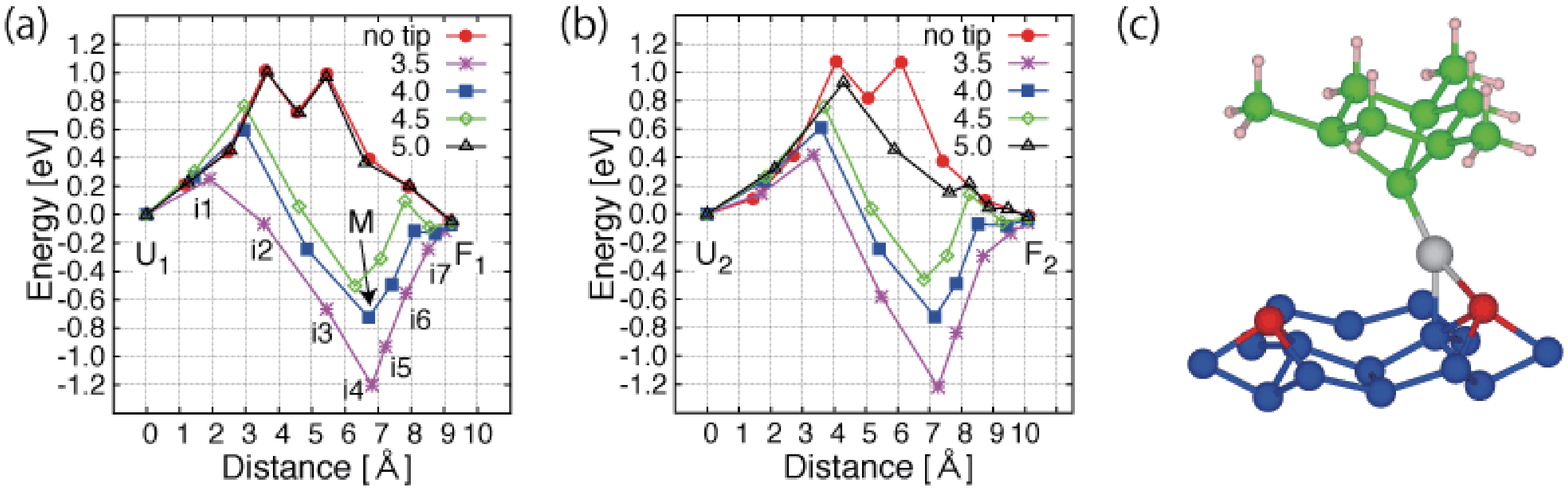}
\caption{
(Color online) The energy profiles of the Ag adatom diffusion between the HUCs on the Si(111) surface and the configuration of the trapped Ag atom. In (a) and (b), the energy profiles along the \textit{pathway 1} and the \textit{pathway 2} with the 4 tip-surface distances are shown, respectively. The circle (red line) is the energy profile of the Ag atom diffusion without the presence of the AFM tip. The triangle (black line), the diamond (green line), the square (blue line) and the asterisk (magenta line) are the energy profiles of the diffusion when the tip-surface distances are 5.0, 4.5, 4.0, and 3.5 \text{\AA}, respectively. The labels and the axes are the same as those in Fig.~\ref{fig:notip}. In (a), the intermediate states between $U_1$ and $F_1$ in case of the tip-surface distance of 3.5 \text{\AA} are labeled by i1 $\sim$ i7. In (c), the enlarged view of the tip-surface atomic configuration of the energy dip labeled as M in (a) is shown. The green, yellow balls are tip Si and H atoms, respectively. The silver ball depicts Ag atom. The red and blue balls are Si adatom, and Si atoms of the lower layers of the surface, respectively.
}
\label{fig:sitip}
\end{figure*}

The diffusion pathways are essentially the same as those of the Ag atom diffusion without the presence of the tip.
The energy profiles along the \textit{pathway 1} and the \textit{pathway 2} with the presence of the Si AFM tip are shown in Fig.~\ref{fig:sitip} (a) and (b), respectively. First, let us describe the diffusion along the \textit{pathway 1}. When the tip-surface distance is 5 \text{\AA}, the energy profile is not modified and is the same as that of the Ag atom diffusion without the AFM tip. However, as the tip further approaches the surface, it is substantially modified. It becomes significantly asymmetric showing that the FHUC $\rightarrow$ UHUC diffusion is less probable than the intended UHUC $\rightarrow$ FHUC diffusion of the Ag atom. There are two kinds of modifications in the energy profile, namely the reduction in the rate determining diffusion barrier and the appearance of local energy minimum. When the tip-surface distance is 4.5 \text{\AA}, the barrier height decreases from 1.01 to 0.77 eV. With the further approach of the tip to the distances of 4.0 and 3.5 \text{\AA}, the barrier height becomes 0.60 and 0.25 eV, respectively. The reduced diffusion barrier indicates that the inter-HUC diffusion is enhanced by the presence of the AFM tip. Unlike the UHUC$\rightarrow$FHUC diffusion, the energy barrier for the diffusion in the reverse direction remains larger than 1 eV at any tip-surface distance, hindering the unintended reverse diffusion to take place. This finding explains the one-way character of the adatom diffusion observed in the experiment.\cite{gate} The drastic decrease of the diffusion barrier is due to the flexibility of the tip apex structure of the Si tip. We explain the situation in detail in the case of the tip-surface distance of 3.5 \text{\AA}. The tip apex atom shifts downward by 0.44 \text{\AA} in the i1 configuration [see Fig.~\ref{fig:sitip} (a)], stretching the backbonds from 2.47 (before structural relaxation) to about 2.64 \text{\AA}. The distance between the tip apex atom and the Ag atom is 4.87 \text{\AA}. The tip apex atom further moves toward the Ag atom in i2 configuration and the 3 backbonds become 2.50, 2.68, and 3.22 \text{\AA} (0.86 \text{\AA} from its rest position). Therefore, the tip reduces the energy barrier of the diffusion from large distance. The importance of the flexibility of the tip apex structure in atom manipulation has been found from our previous work.\cite{bato} The current finding is a corroboration of such general statement.

The dip (or the local energy minimum) in the energy profile also deepens as the tip-surface distance becomes small. The large dip in the energy profile shows that the Ag atom is trapped by the AFM tip proving the picture provided by the experimenters correct.\cite{gate} However, in the configuration corresponding to the energy dip, the Ag atom is not right under the tip apex atom. Rather, it is bonding with multiple Si atoms in agreement with the previous study that stated the preference of Ag atom for the multi-coordination with the Si atoms.\cite{agondastheory} As an example, the atomic structure of the M-configuration [see Fig.~\ref{fig:sitip} (a)] is shown in Fig.~\ref{fig:sitip} (c). The Ag atom is bonding with 3 Si atoms: tip apex atom, Si adatom (red ball), and Si atom in lower layer. The bond length between the Si adatom and the lower layer Si atom, bonding with the Ag atom, is stretched from 2.47 to 2.68 \text{\AA}. When the AFM tip approaches further, the Ag atom starts to interact with the second layer Si atoms of the tip. In the i4 configuration with the tip-surface distance of 3.5 \text{\AA}, the Ag atom bonds with 4 Si atoms: 2 Si atoms of the tip and 2 Si atoms of the surface.

The modifications of the energy profile of the diffusion along the \textit{pathway 2} are essentially the same as those of the \textit{pathway 1}. The barrier heights become 0.93, 0.76, 0.61, and 0.42 eV when the tip-surface distances are 5.0, 4.5, 4.0, and 3.5 \text{\AA}, respectively. The Ag atom is also trapped by the tip when the tip-surface distance is small enough. The atomic configurations of the trapped states are similar to those of the \textit{pathway 1}, in which the Ag atom bonds with 3 or 4 Si atoms. Such modifications are due to the bond formation between the tip apex atom and the Ag atom. It is confirmed by the charge density distribution analysis. The analysis shows that the charge redistributes and accumulates in the intermediate region between the Ag atom and the tip apex Si atom when the distance between them are short.

\begin{figure}
\centering
\includegraphics[width=0.8\linewidth]{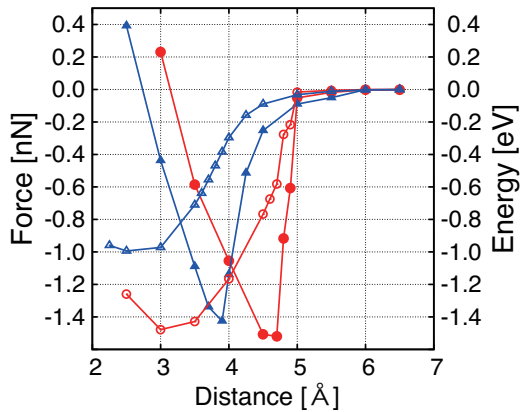}
\caption{
(Color online) The force-distance and energy-distance curves of Si and Pt tips over the Ag atom on the Si(111) surface. The x-axis is the distance between tip apex atom and the Ag atom on the Si surface before structural relaxation. The left and right y-axes are the force acting on the AFM tip and the energy gain in the system due to the interaction of the tip and the surface, respectively (see text). The blue line with the filled and unfilled triangles are the force-distance and energy-distance curves of the Pt tip, respectively. The red line with the filled and unfilled circles are the force-distance and energy-distance curves of the Si tip, respectively. The negative values of force and energy mean attractive force on the tip and energy gain in the total energy, respectively.
}
\label{fig:force}
\end{figure}

The energy profiles of the diffusion along the \textit{pathway 1} and the \textit{pathway 2} when the tip-surface distance is 5.0 \text{\AA} are significantly different. The energy profile of the diffusion along the \textit{pathway 2} becomes asymmetric, while that of the diffusion along the \textit{pathway 1} remains almost unchanged by the tip. It is due to the nature of the interaction between the tip apex atom and the Ag atom. To get an insight to this, we consider a simple case in which the Si tip approaches the Ag atom adsorbed at $U_1$ site from above. The Ag atom is at the site $U_1$ for it is the most stable adsorption site in the HUC. The obtained force acting on the tip and the change in the total energy of the system with the decreasing distance between the tip apex and Ag atom are shown in Fig.~\ref{fig:force}. As shown in the force-distance curve of Si tip placed over the Ag atom, at around 5.0 \text{\AA}, slight change in the distance between the tip apex atom and the Ag atom results in sudden increase in the force acting on the AFM tip. Therefore, the difference in the energy profiles shows that the Ag atom in the \textit{pathway 1} diffuses closer to the tip apex atom than the Ag atom in the \textit{pathway 2}.

\subsubsection{\label{subsubsec:pt_tip} The effect of the Pt tip on the diffusion}

\begin{figure}
\centering
\includegraphics[width=0.7\linewidth]{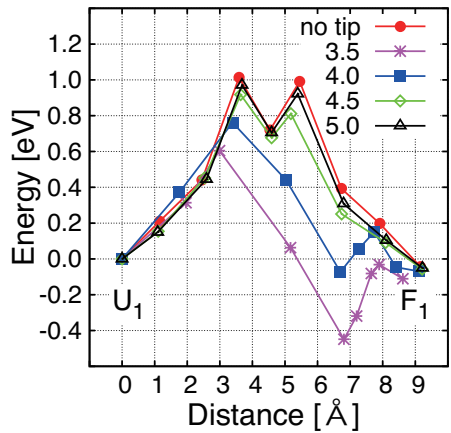}
\caption{
(Color online) 
The energy profiles of the Ag atom diffusion with the presence of the Pt tip. The labels and the axes are the same as those in the Fig.~\ref{fig:sitip}.
}
\label{fig:pttip}
\end{figure}

Here we present the modifications by the Pt tip of the diffusion of the Ag atom on the Si(111) surface. The energy profiles of the Ag atom diffusion from $U_1$ to $F_1$ along the \textit{pathway 1} with the different tip-surface distances are shown in Fig.~\ref{fig:pttip}. With the shorter tip-surface distance, the energy barrier of the Ag atom diffusion is reduced and the energy profile becomes asymmetric with the appearance of the local energy minimum. However, as it can be seen, the Pt tip does not decrease the energy barrier as drastically as the Si tip. With the tip-surface distance of 5.0, 4.5, 4.0, and 3.5 \text{\AA}, the energy barrier of the rate determining process becomes 0.97, 0.92, 0.72 and 0.60 eV, respectively. These energy barriers are significantly higher than the energy barriers modified by the Si tip. This difference between the Si and Pt tips are due to the difference in the flexibility of the tip apex structure. Unlike the Si tip, the tip apex atom of the Pt tip does not move much to create bond with the Ag atom. During the diffusion, the Pt atom shifts 0.19 \text{\AA} at most from its unrelaxed position when the tip surface distance is 3.5 \text{\AA}. The force-distance curve and the change in the total energy of the system, in which Pt tip approaches the Ag atom from above, are shown in Fig.~\ref{fig:force}. Considering that the maximum attractive forces acting on the Si and Pt tips are almost the same when the tip is interacting with the Ag atom on the Si surface (see Fig.~\ref{fig:force}) and the difference in the shift of the tip apex atoms of Si and Pt tips, we conclude that the Pt tip is stiffer than the Si tip.

The depth of the energy dip is shallower than that of the Si tip case with the same tip-surface distance. This is due to the difference between the Si-Ag and Pt-Ag bonds, and the sharpness and the stiffness of the tip apex structure. As mentioned above, the Pt tip apex atom does not move toward the Ag atom. Therefore the bond length of the tip apex atom and the Ag atom bond stays longer than that of Si tip. Moreover, as plotted in Fig.~\ref{fig:force}, the energy gain by the formation of the Pt-Ag bond is about 1.0 eV at most which is much smaller than that of the 1.6 eV of the Si-Ag bond. The considered Pt tip is sharper than the Si tip. Even at the 3.5 \text{\AA} tip-surface distance, the Ag atom is interacting with the tip only by the tip apex Pt atom unlike the Si tip case. But with the smaller tip-surface distance, it is expected that the Ag atom would interact with the second layer atom of the tip and the trapping effect would be enhanced.

Even though less substantial than that by the Si tip, our results show that the diffusion controlling of the Ag atom diffusion between HUC of Si(111) surface is possible with Pt tip. This is consistent with the experimental situation in which Pt-Ir coated tip is successfully utilized to control the inter-HUC diffusion.\cite{gate}

\section{\label{sec:summary} Summary}

We have performed total-energy electronic-structure calculations using density functional theory for the diffusion of the Ag atom between the half unit cells (HUCs) on the Si(111)-(7$\times$7) surface with and without the probing tip of the atomic force microscope (AFM). We have first clarified the atom-scale reaction pathways and the corresponding energy barriers for the Ag atom diffusion on the surface. There are 2 pathways of the inter-HUC diffusion connecting the (meta)stable adsorption sites in the two HUCs. In the middle of each pathway, there is a metastable adsorption site rendering the diffusion to be a two-step process. The energy barriers of the rate determining processes in unfaulted-HUC to faulted-HUC diffusion (forward diffusion) and the reverse diffusion (backward diffusion) are about 1 eV, indicating that the inter-HUC diffusion is rare event at room temperature. We have also identified the reaction pathways and the corresponding energy barriers for the diffusion with the presence of the Si and Pt tips of the AFM. When the tip is placed slightly off to the side of the faulted-HUC from the boundary of the HUCs, the energy barriers of the forward and the backward diffusions become asymmetric. For both pathways, the energy barriers of the intended diffusion (forward diffusion) decrease with the small tip-surface distance. The Si tip reduces the barrier more substantially than the Pt tip. We have found that the flexibility of the tip apex structure was crucial in the drastic lowering of the energy barrier. In addition to the barrier lowering effect, the tip traps the adatom near the tip apex, preventing the backward diffusion to take place. The energy barrier of the backward diffusion stays larger than 1 eV for any tip-surface distance. The trapping effect of the tip is enhanced by the interaction between the diffusing adatom and the second layer atom of the tip. Therefore, the tip with the blunt tip apex structure has larger trapping ability than the tip with sharper tip apex structure. The bond formation between the AFM tip atom and the surface adatom is the essential physics for the atom manipulation. Our calculations show that the diffusion controlling is possible with both metallic and semiconducting AFM tips.

\begin{acknowledgments}
This work was supported by MEXT as a social and scientific priority issue (Creation of new functional devices and high-performance materials to support next-generation industries) to be tackled by using post-K computer. Computations were performed mainly at the Supercomputer Center at the Institute for Solid State Physics, The University of Tokyo, The Research Center for Computational Science, National Institutes of Natural Sciences, and the Center for Computational Science, University of Tsukuba.
\end{acknowledgments}

\end{document}